\documentclass{cimento2}

\usepackage{amsmath}
\usepackage{latexsym}
\usepackage{amsfonts}
\usepackage{float}
\usepackage{bm}
\usepackage{mathabx}
\usepackage{amsbsy}
\usepackage{amssymb}
\usepackage{epsf}
\usepackage{graphicx}

\newcommand{\be}{\begin{equation}}\newcommand{\ee}{\end{equation}}
\newcommand{\bea}{\begin{eqnarray}}\newcommand{\eea}{\end{eqnarray}}
\newcommand{\brr}{\begin{array}}\newcommand{\err}{\end{array}}
\newcommand{\bit}{\begin{itemize}}\newcommand{\eit}{\end{itemize}}
\newcommand{\ben}{\begin{enumerate}}\newcommand{\een}{\end{enumerate}}

\newcommand{\bbm}{\begin{bmatrix}}\newcommand{\ebm}{\end{bmatrix}}

\newcommand{\ba}{\begin{array}}
\newcommand{\ea}{\end{array}}
\newcommand{\G}{\textbf}

\newtheorem{mydef}{Definition}
\newtheorem{Lemma}{Lemma}
\newtheorem{theorem}{Theorem}
\newcommand{\bd}{\begin{mydef}} \newcommand{\ed}{\end{mydef}}
\newcommand{\bthe}{\begin{theorem}} \newcommand{\ethe}{\end{theorem}}
\newcommand{\ble}{\begin{Lemma}} \newcommand{\ele}{\end{Lemma}}

\def\ha{\frac{1}{2}}

\def\tr{\mathrm{tr}}

\def\lf{\left}
\def\noi{\noindent}
\def\non{\nonumber}\def\ran{\rangle}
\def\rar{\rightarrow}
\def\ri{\right}
\def\al{\alpha}\def\bt{\beta}\def\ga{\gamma}
\def\de{\delta}
\def\te{\vartheta}
\def\la{\lambda}\def\si{\sigma}
\def\om{\omega}

\def\1{{_{1}}}\def\2{{_{2}}}

\newcommand{\ide}{1\hspace{-1mm}{\rm I}}

\def\noHe0{:\;\!\!\;\!\!:H_e(0):\;\!\!\;\!\!:}
\def\noHm0{:\;\!\!\;\!\!:H_\mu(0):\;\!\!\;\!\!:}

\def\vect#1{{\bm #1}}

\def\lf{\left}
\def\noi{\noindent}
\def\non{\nonumber}
\def\ran{\rangle}
\def\rar{\rightarrow}
\def\ri{\right}

\def\al{\alpha}\def\bt{\beta}\def\ga{\gamma}
\def\de{\delta}
\def\te{\theta}
\def\la{\lambda}
\def\si{\sigma}
\def\om{\omega}

\def\1{{_{1}}}\def\2{{_{2}}}

\def\I{{_{\rm{I}}}}\def\II{{_{\rm{II}}}}
\def\A{{_{A}}}\def\B{{_{B}}}

\title{Effective action approach to dynamical generation of fermion mixing}
\author{M.~Blasone\from{InstA}\from{InstB},
P.~Jizba\from{InstC} and
L.~Smaldone\from{InstA}\from{InstB}}
\instlist{\inst{InstA} Dipartimento di Fisica, Universit\`a di Salerno -- Fisciano, Italy
\inst{InstB} INFN Sezione di Napoli, Gruppo collegato di Salerno, Italy
  \inst{InstC} FNSPE, Czech Technical University in Prague -- Prague, Czech Repulic}

\PACSes{\PACSit{12.15.Ff}{Quark and lepton masses and mixing}
\PACSit{03.70.+k}{Theory of quantized fields}}

\begin{document}

\maketitle

\begin{abstract}
In this paper we discuss a mechanism for the dynamical generation of flavor mixing, in the
framework of the Nambu--Jona Lasinio model. Our approach is illustrated both with the conventional operatorial formalism
and with functional integral and ensuing one-loop effective action. The results obtained are briefly discussed.
\end{abstract}

\section{Introduction}

The phenomenon of mixing of fields subject to different interactions is of considerable importance in current particle-physics phenomenology.
This is particularly the case for
flavor oscillations of neutrinos or  oscillations of strangeness or beauty in certain mesons (e.g., $K^0-\bar{K}^0$ or $B^0-\bar{B}^0$ mixing, where the
mixing is also closely related to the mechanism of CP violation).
It has been observed in recent years~\cite{Blasone:1995zc} that the usual mixing transformations, which are simple rotations at  level of quantum fields, do indeed contain  Bogoliubov transformations at level of annihilation/creation operators. This key observation directly implies that the vacuum for  neutrino fields with definite flavor, the flavor vacuum, has a non trivial  structure of a condensate of  particle/antiparticle pairs. Starting from this initial insight, a number of consequences have been
derived, including corrections to the standard quantum mechanical Pontecorvo oscillation formulas~\cite{BHV98}.

In this context a natural question arises whether the above vacuum structure could not be seen as the result of a dynamical mechanism for the mixing generation~\cite{sarkar},
in some sense akin to mechanism that is responsible for  dynamical mass generation in Nambu--Jona Lasinio (NJL) model~\cite{NJL}. While pondering this possibility, one should not
ignore a remarkable formulation introduced in 1964 by Umezawa, Takahashi and Kamefuchi~\cite{UTK}. There the authors established a connection between the inequivalent representations
of the commutation relations and the mechanism of dynamical breaking of symmetry and dynamical generation of the fermion mass. They studied, in particular, the NJL model and
the associated gap equation. By following this formalism, it has been recently demonstrated~\cite{DynMix} that in the case of a two-flavors NJL model,
mixing terms arise in connection with particular inequivalent representations, which are not usually taken into account.

In this paper, we extend the above (purely operatorial) analysis in terms of functional integrals and associated one-loop effective action technique. The results of our study seem to confirm our previous findings~\cite{DynMix}
and lead in a straightforward way to the gap equations which account for the generation of both masses and mixing terms in the theory.

\section{Dynamical generation of mass in the UTK formalism}

Following Ref.\cite{UTK}, we consider a system of {\em Fermi} fields enclosed in a finite-volume (volume $V$) box.
Let  $|0\rangle$ be a fiducial reference vacuum state with the corresponding set of annihilation operators for particles and antiparticles,
$a_{\bf{k}}^r$ and $b_{\bf{k}}^r$, satisfying the anticommutation relations
\begin{eqnarray}
\lf\{a_{\bf{k}}^r, a_{\bf{l}}^{s\dag}\ri\} \ = \ \lf\{b_{\bf{k}}^r, b_{\bf{l}}^{s\dag}\ri\} \ = \ \delta_{\bf{k},\bf{l}}\delta_{rs}\, ,
\end{eqnarray}
with other anticommutators being zero. Here $r=1,2$ is the helicity index and
\begin{eqnarray}
{\bf k} \ = \ \frac{2\pi}{V^{1/3}}\ \!{\bf{n}}, \;\;\;\;\; n_1, n_2, n_3 \;\;\;{\mbox{integers}}\, .
\end{eqnarray}
The plane-wave expansion for the field is:
\begin{eqnarray}\label{freefield}
&& \psi(x) \ = \ \frac{1}{\sqrt{V}} \sum_{{\bf k},r } \left[u_{{\bf k}}^r \ a_{{\bf k}}^r \ \!e^{  i{\bf k}\cdot{\bf x}}
 +v_{{\bf k}}^r \ b_{{\bf k}}^{r\dag} \ \!e^{  -i{\bf k}\cdot {\bf x}}\right],
 \end{eqnarray}
where the spinor wavefunctions $u_{{\bf k}}^r$, $v_{{\bf k}}^r$ carry the time dependence through the factors $e^{- i \om_k t}$ and $e^{ i \om_k t}$, respectively, with $\om_{\bf k}=\sqrt{{\bf k}^2 + m^2}$.

In quantum field theory (QFT) the Hilbert space is not uniquely defined: the infinite number of degrees of freedom allows for the existence of unitarily inequivalent representations of
the canonical (anti)-commutation relations (CAR)~\cite{BJV}-\cite{Miransky}. This fact is at the heart of the symmetry breaking mechanism~\cite{BJV,Umezawa:1982nv}. On the other hand, in QM, where the volume is taken to be finite and number
of particles is fixed, all the representations of the CAR are unitarily (i.e., physically) equivalent. Thus, in order to formulate the QFT Hilbert-space structure we start with a regulated finite-volume space and look for
unitary transformations of the vacuum state $|0\rangle$ that satisfy simple (physically motivated) consistency criteria. Then, in the large volume limit, we construct all possible candidates physical Fock spaces.

The generator of such (finite volume) unitary transformations $G$ can be parametrized with only two parameters $\vartheta_{{\bf k}}^{r}$ and $\varphi_{{\bf k}}^{r}$, namely
\begin{eqnarray}
G(\vartheta,\varphi) \ = \ \exp\left[\sum_{{\bf k},{r}} \vartheta_{{\bf k}}^{r}\ \!\left(b_{-\bf{k}}^{r}a_{\bf{k}}^r
e^{-i\varphi_{{\bf k}}^{r}} -
a_{\bf{k}}^{r\dag} b_{-\bf{k}}^{r \dag }e^{i\varphi_{{\bf k}}^{r}} \right)   \right].
\end{eqnarray}
The invariance of $G$ under rotation ensures that $\vartheta$ and $\varphi$ depend only on $k\equiv |{\bf k}|$. In addition, it can be argued~\cite{UTK} that $\vartheta$ is independent on $r$ and $\varphi_k^r=- (-1)^r \varphi_k$. The transformations generated by $G$ are
\begin{eqnarray}
\alpha_{{\bf k}}^r  \ &=& \ G(\vartheta,\varphi) a_{\bf{k}}^r G^{\dag}(\vartheta,\varphi)
= \ \cos\vartheta_k \ \! a_{\bf{k}}^r + e^{i\varphi_k^{r}} \sin \vartheta_k
\ \! b_{-\bf{k}}^{r \dag }\, , \label{II.8.a}  \\[3mm]
\beta_{{\bf k}}^r  \ &=& \ G(\vartheta,\varphi)b_{\bf{k}}^r G^{\dag}(\vartheta,\varphi)
= \ \cos \vartheta_k \ \! b_{\bf{k}}^r -
e^{i\varphi_k^{r}} \sin \vartheta_k  \ \! a_{-\bf{k}}^{r\dag}\, .\label{II.8.b}
\end{eqnarray}

These are Bogoliubov transformations,  preserving the canonical commutation relations.
The labels $\{\vartheta_k,\varphi_k^{r}\}$ yield the most general  parametrization for the Bogoliubov transformation of creation and annihilation operators. The vacuum state for the $\alpha_{{\bf k}}^{r}$ and $\beta_{{\bf k}}^{r}$ operator is given by
\begin{eqnarray}
|0(\vartheta,\varphi)\rangle \ = \ G(\vartheta,\varphi) |0\rangle  \ = \ \prod_{{\bf k},{r}}\left(\cos\vartheta_k \ - \ e^{i\varphi_k^{r}}
\sin \vartheta_k \ \! a_{\bf{k}}^{r\dagger} b_{-\bf{k}}^{r\dagger}   \right)\!|0\rangle\,.
\label{II.10.aa}
\end{eqnarray}

In the finite volume limit all vacuum states $|0(\vartheta,\varphi)\rangle$ are   equivalent
(i.e., they describe the same unique physical ground-state).
In the infinite-volume limit the situation is drastically different.  This can be seen by noticing that from (\ref{II.10.aa}) we have (for $V\rar \infty$):
\begin{eqnarray}
&&\langle 0| 0(\vartheta,\varphi)\rangle \ = \ \exp\left[\sum_{{\bf k}, r} \log(\sin \vartheta_k) \right] \ = \ \exp\left[\frac{V}{(2\pi)^3} \int d^3 {\bf k} \log(\sin^2 \vartheta_k) \right] \rightarrow \ 0\, .
\end{eqnarray}
More generally, in the infinite-volume limit all the physical vacua with different $\vartheta$'s and $\varphi$'s are orthogonal, i.e.,
\begin{eqnarray}
\langle  0(\vartheta,\varphi)| 0(\vartheta',\varphi') \rangle \  \rightarrow \  0\, , \;\;\;\;\;\; \vartheta',\varphi' \neq\ \vartheta,
\varphi\, .
\end{eqnarray}

The free field (\ref{freefield}) can be expressed in the representation $\{\vartheta,\varphi \}$ by
 means of the Bogoliubov transformation (\ref{II.8.a})-(\ref{II.8.b}):
\begin{eqnarray}
&&
\psi (x) \ = \ \frac{1}{\sqrt{V}} \sum_{{\bf k},r } \left[u_{{\bf k}}^r(\vartheta,\varphi) \alpha_{{\bf k}}^r \ \!e^{  i{\bf k}\cdot{\bf x}}
 +v_{{\bf k}}^r(\vartheta,\varphi) \beta_{{\bf k}}^{r\dag} \ \!e^{  -i{\bf k}\cdot {\bf x}}\right], \label{psiexex}
 \end{eqnarray}
with $ \alpha_{{\bf k}}^r\, |0(\vartheta,\varphi)\rangle \ = \ \beta_{{\bf k}}^{r}\,  |0(\vartheta,\varphi)\rangle \ = \ 0\, . $
The Dirac spinors $u_{{\bf k}}^r(\vartheta,\varphi)$ and $v_{{\bf k}}^r(\vartheta,\varphi)$ are related with
the fiducial representation spinors via the relations
\begin{eqnarray}
&&u_{{\bf k}}^r(\vartheta,\varphi) \ = \ u_{{\bf k}}^r \ \!  \cos \vartheta_k \ + \ v_{-{\bf k}}^r \ \! e^{-i\varphi_k^{r}} \sin \vartheta_k\, , \\[2mm]
&&v_{{\bf k}}^r(\vartheta,\varphi) \ = \ v_{{\bf k}}^r \ \! \cos \vartheta_k \ - \ u_{-{\bf k}}^r \ \! e^{i\varphi_k^{r}}\sin \vartheta_k\, .
\end{eqnarray}
It is to be remarked that the expressions (\ref{psiexex}) and  (\ref{freefield}) represent
indeed the same operator, expanded in terms of different sets of creation/annihilation operators which
act on different (orthogonal) vacua.

Let us then consider the so-called $V$-limit procedure introduced by Umezawa {\em et al.} in Ref.~\cite{UTK}.
One takes
matrix elements of QFT operators, say $Q$, between states $|\Phi_i(\vartheta,\varphi)\rangle$, generated from the vacuum state $|0(\vartheta,\varphi)\rangle$
by a suitable action of creation and annihilation operators. The index ``$i$" is a multi-index distinguishing various states, and the two real parameters $\vartheta$ and $\varphi$ label the different (unitarily inequivalent) vacuum states.
In particular the $V$-limit of $Q$ with respect to a representation characterized by the parameters $\{\vartheta,\varphi\}$
is defined as
\begin{eqnarray}
&&\langle \Phi_i(\vartheta,\varphi)| \mbox{$V$-lim}[Q]|\Phi_j(\vartheta,\varphi)\rangle \equiv \ \lim_{V\rightarrow \infty}\langle \Phi_i(\vartheta,\varphi)| Q|\Phi_j(\vartheta,\varphi)\rangle\, ,
\label{Eq.5a}
\end{eqnarray}
for all $i$ and $j$. The matrix element on the right-hand side of (\ref{Eq.5a}) is operationally calculated by phrasing the full (Heisenberg-picture) fields $\psi$ present in $Q$ in terms of the asymptotic fields $\psi_{\rm in}$ enclosed in a finite-volume (volume $V$) box. The mapping between $\psi$ and $\psi_{\rm in}$  is the Yang--Feldman equation (see also~\cite{BJV,Yang-Feldman,BJ02AP}). Formally it can be written in the form~\cite{BJV}; $\psi(x) = S^\dag T(S \psi_{\rm in}(x))$, where $S$ and $T$ are the $S$-matrix and time-ordering symbol, respectively.

Thank this result we can calculate the following useful quantities:
\vspace{-3mm}
\bea\non
&&	C_p\equiv i \lim_{V\rightarrow \infty}\,
\langle 0(\vartheta,\varphi)| \bar{\psi}(x) \ga_5 \psi(x)|0(\vartheta,\varphi)\rangle \
=\ \frac{2}{(2 \pi)^3} \int d^3{\bf k}\, \sin 2\vartheta_k \ \sin \varphi_k
\\                      && C_s\equiv \lim_{V\rightarrow \infty}
\langle 0(\vartheta,\varphi)| \bar{\psi}(x)  \psi(x)|0(\vartheta,\varphi)\rangle\ =\ - \frac{2}{(2 \pi)^3} \int d^3{\bf k}\, \lf[ \frac{m}{\om_k} \cos 2\vartheta_k \,
 - \frac{k}{\om_k} \sin 2\vartheta_k \cos \varphi_k \ri]. \non
 \\  \label{CpCs}
\eea
We now consider the dynamical mass generation in NJL model for the case of one flavor.  Here we shall follow closely the simplified presentation given in \cite{DynMix}, with the aim to expose the main logical passages of treatment given in Ref.\cite{UTK}.  The NJL is described by the following Hamiltonian
\begin{eqnarray}
                          &&{H} \ = \ {H}_0 + {H}_{\rm{int}}\, , \\[2mm]
&&{H}_0 \ = \ \int d^3{\bf{x}} \ \! \widebar{{{\psi}}}  \left( -i\vect{\gamma}\cdot\!\vect{\nabla}\ + \  m\right) {{\psi}}\, ,
  \\[1mm]
&&{H}_{\rm{int}} \ = \ \lambda \int d^3 {\bf{x}} \ \! \left[\left(\widebar{{{\psi}}} {{\psi}}\right)^2  -
\left(\widebar{{{\psi}}} \gamma^5{{\psi}}\right)^2\right]\, .
\label{Eq.3.1}
                      \end{eqnarray}
We take in general  $m\neq 0$, the case $m=0$ is then obtained as a special case.	
										
Considering the lowest order in the Yang--Feldman expansion, the $V$-limit of $H$ gives:
\begin{eqnarray}
 && V\mbox{-lim}\left[{H}\right] \ = \  \widebar{ H}_0
 \,  + \, \mbox{c-number}\, ,
\\  \label{VlimH}
&& \widebar{H}_0\, =\, H_0 + \delta H_0\, , \qquad
\delta H_0 \  =\  \int d^3{\bf{x}} \ \lf( f \, \widebar{\psi} \psi  \ + \ i  g \, \widebar{\psi} \ga_5 \psi \ri).
\end{eqnarray}
with $
f \, = \, \la\, C_s $ , $ g \, = \, \la\, C_p\, .$
																						
Until now the state $|0(\vartheta,\varphi)\rangle$ is not specified: this choice fixes the representation and has to be done on physical basis. It is then required \cite{UTK}  that  the $V$-limit of the full Hamiltonian $H$
should describe the quasiparticle  (free) Hamiltonian with the correct relativistic dispersion relation, namely:
\begin{eqnarray} \label{H0diag}
\widebar{H}_0  \ = \ \sum_{r} E_{k}\left(\alpha_{{\bf k}}^{r\dag}\alpha_{{\bf k}}^r +
\beta_{{\bf k}}^{r\dag}\beta_{{\bf k}}^r \right) \ + \ W_{0}\, ,
\end{eqnarray}
with   $E_k = \sqrt{k^2 + M^2}$. The mass $M$ corresponds to the mass of elementary excitations
(or quasiparticles) over the physical vacuum. $W_0$ is the vacuum energy (or condensate density) and is given \cite{UTK} by $ W_0=-2 \int d^3{\bf{k}} \, E_k $.

One finds that the condition \eqref{H0diag} is satisfied\footnote{The condition $E_k>0$ is also enforced.} when the
following conditions hold
\begin{eqnarray}
\cos(2\vartheta_k) & = &
\frac{1}{E_{k}}\left[\frac{m}{\omega_{k}} f(\vartheta, \varphi) +
\omega_{k}\right], \label{MGE1a} \\[2mm]
\sin(\varphi_{k}^r) & = & g(\vartheta, \varphi)(-1)^r  \ \!\left[g^2(\vartheta, \varphi) +
\frac{{k}^2}{\omega_{k}^2} f^2(\vartheta, \varphi)\right]^{-\frac{1}{2}}, \label{MGE1b}  \\[2mm]
M^2(\vartheta, \varphi)  & = & (m \ + \ f(\vartheta, \varphi))^2 + g^2(\vartheta, \varphi)
 = \  (m \ + \ \la C_s)^2 + \la^2 C_p^{\,2}\, ,
\label{MGE1}
\end{eqnarray}
Since $f$ and $g$ depend on the parameters $\{\vartheta, \varphi\}$, the above solutions give rise to  two non-linear equations
\begin{eqnarray}
f \ = \ f(\vartheta(f,g), \varphi(f,g))\;\;\;\; \mbox{and} \;\;\;\; g \ = \ g(\vartheta(f,g), \varphi(f,g))\, ,
\label{eq-39}
   \end{eqnarray}
which can be recasted as
\bea
C_p \lf(1 + \frac{2\la}{(2\pi)^3}\int \frac{d^3{\bf k}}{E_k}\ri) &=&0\,,
\\
C_s \lf(1 + \frac{2\la}{(2\pi)^3}\int \frac{d^3{\bf k}}{E_k}\ri) &=& - \frac{2 m}{(2\pi)^3}\int \frac{d^3{\bf k}}{E_k}\,.
\eea
These equations determine the mass M.
In \cite{UTK} two possibilities are discussed
\bea \label{gapeqA}
C_p = 0, & & M = m \,-\, \frac{2\la }{(2\pi)^3}\, M \int \frac{d^3{\bf k}}{E_k},
\\ \label{gapeqB}
m = 0, & & 1 + \frac{2\la}{(2\pi)^3}\int \frac{d^3{\bf k}}{E_k} \, =\,0.
\eea
The second case, Eq.~\eqref{gapeqB}, is only allowed for $\la <0 $ and for $m=0$ is nothing but the gap equation!
For $m\neq 0$, Eq.~\eqref{gapeqA} gives perturbative corrections to the mass:
\bea
M &=& m \, - \, \frac{2\la }{(2\pi)^3}\, m \int \frac{d^3{\bf k}}{\om_k} \, + \, \cdots
\eea
On the other hand, the solution Eq.~\eqref{gapeqB} has a non-perturbative character and expresses the dynamical breakdown of (chiral) symmetry.

\section{Dynamical generation of flavor mixing - Operatorial approach}

We now consider the dynamical symmetry breaking for the case of two fermion fields, for which in general a non-diagonal mass matrix will be obtained, thus generating flavor mixing in addition to nonzero masses. We expose first the operatorial approach, as analyzed in \cite{DynMix}. Here the notation is over-simplified,  spacetime dependence is omitted as well as momentum and helicity indices. Let us consider a fermion field doublet ${\vect{\psi}}$ whose Hamiltonian density is given as
\begin{eqnarray}\label{generalH}
&&{\cal H} \ = \ {\cal H}_0 + {\cal H}_{\rm{int}}\, ,\\[2mm]
&&{\cal H}_0 \ = \  {\widebar{\vect{\psi}}} \left(
-i\vect{\gamma}\cdot\!\vect{\nabla}\ + \ \mathrm{\G M}_0\right)
 {\vect{\psi}}\, ,
\label{Eq.1}
\end{eqnarray}
with $\vect{\gamma}$ being a shorthand for $\ide\otimes\vect{\gamma}$ with  $\ide$ being the $2\times 2$ identity matrix and
\begin{eqnarray}
{\vect{\psi}} \ = \ \left(
\begin{array}{c}
 \psi_{_{\rm{I}}} \\
 \psi_{_{\rm{II}}} \\
 \end{array}
\right) \;\;\;\;\; \mbox{and} \;\;\;\;\; \mathrm{\G M}_0 \ =
 \ \left( \begin{array}{cc}m_{_{\rm{I}}} & {0}\\{0}
& m_{_{\rm{II}}}
 \\ \end{array} \right) \, .
\label{Eq.2}
\end{eqnarray}
The interaction Hamiltonian ${H}_{\rm{int}}$ can be assumed in the generic form
\begin{eqnarray}\label{Hintfull}
{\cal H}_{\rm{int}} \ = \
\left(\widebar{{\vect{\psi}}}\,\Gamma \,
{\vect{\psi}}\right)\left(\widebar{{\vect{\psi}}}\, \Gamma' \, {\vect{\psi}}\right),
\end{eqnarray}
where $\Gamma$ and $\Gamma'$ are some doublet spinor matrices. For simplicity in this case we consider only the scalar counterterms, i.e., we set $g_\I = g_\II =0$. This in turn implies that $\varphi_\I=\varphi_\II=0$ in the Bogoliubov transformations for fields $\psi_\I$ and $\psi_\II$. This assumption simplifies considerably the
following treatment, without altering the main results of our analysis.

The term $\delta \mathcal{H}_0$ arising from $V$-limit has now generally the  following
form
\begin{eqnarray} \non
\delta \mathcal{H}_0  &=&  \delta \mathcal{H}_0^\I \, +\, \delta
\mathcal{H}_0^\II\, +\,
\delta \mathcal{H}_{\rm{mix}}\\[2mm]
&=& f_\I \,\widebar{\psi}_\I \psi_\I
\ + \ f_\II\, \widebar{\psi}_\II \psi_\II \ + \
 h  \, \left(\widebar{\psi}_\I \psi_\II \ + \
{\widebar \psi}_\II \psi_\I \right)\, .
\end{eqnarray}
Instead of the  Bogoliubov transformations Eqs.\eqref{II.8.a}-\eqref{II.8.b},
we have now a more general $4\times 4$ canonical transformation, defining inequivalent representations. This can be conveniently
parametrized as:
\bea
\bbm \alpha_A \\ \beta_A^\dagger \\ \alpha_{B}\\ \beta_{B}^\dagger \ebm
= \bbm
c_\theta\, \rho_{\A\I}& c_\theta \,\lambda_{\A\I} &
s_\theta \,\rho_{\A\II}  &
s_\theta \,\lambda_{\A\II}
\\  - c_\theta \,\lambda_{\A\I} &c_\theta\, \rho_{\A\I} & - s_\theta
\,\lambda_{\A\II} & s_\theta \,\rho_{\A\II}
\\  - s_\theta \,\rho_{\B \I} &-  s_\theta \,\lambda_{\B \I}& c_\theta
\,\rho_{\B \II}
&  c_\theta \,\lambda_{\B \II} \\   s_\theta \,\lambda_{\B \I} &-
s_\theta\,
\rho_{\B \I} & - c_\theta\, \lambda_{\B \II}& c_\theta\, \rho_{\B \II}
\ebm
  \bbm a_\I \\ b_\I^\dagger \\  a_\II
\\b_\II^\dagger \ebm.
  \label{4x4Bog}
\eea
where $c_\theta\equiv \cos\theta$, $s_\theta\equiv \sin\theta$ and
\begin{eqnarray}\label{rholambda}
  \rho_{a b} \ \equiv \ \cos\frac{\chi_a - \chi_b}{2}, \quad
\lambda_{a b} \ \equiv \  \sin\frac{\chi_a -
\chi_b}{2} \, ,
\quad
\chi_a \ \equiv \  \cot^{-1}\lf[\frac{k}{m_a}\ri]\, ,
\end{eqnarray}
with $a,b={\rm I},{\rm II},A,B$. The transformation~\eqref{4x4Bog} contains  three parameters $(\theta, m_A, m_B)$ to be fixed in terms of the quantities ($f_\I,f_\II,h$) on the basis of physical considerations.

 We first consider the case in which no mixing arises after the $V$-limit. Then the Hamiltonian reduces into the sum of two  Hamiltonians, each of the form as in Eq.(\ref{VlimH}):
\be
  {\widebar {\cal H}}_0 \ = \ \sum_{i={\rm I},{\rm II}}\lf({\cal H}_0^i  \ + \  \delta \mathcal{H}_0^i  \ri).
  \label{48a}
\ee
															
In this case, the Bogoliubov matrix $\eqref{4x4Bog}$ becomes block diagonal:
\bea
  \bbm \alpha_A \\ \beta_A^\dagger
\\
\alpha_{B} \\ \beta_{B}^\dagger \ebm
\ = \
\bbm   \rho_{\A\I}&  \lambda_{\A\I} &
0 &0
\\- \lambda_{\A\I} &  \rho_{\A\I} &  0 & 0
\\ 0 &0& \rho_{\B \II}
&  \lambda_{\B \II} \\ 0 &0 &   -\lambda_{\B \II}& \rho_{\B \II}
\ebm
\bbm a_\I \\ b_\I^\dagger \\ a_\II
\\b_\II^\dagger \ebm
\eea
and the diagonalization condition reads (cf Eq.\eqref{MGE1}):
\bea
  m_A \ = \ m_\I+f_\I\,, \quad m_B \ = \  m_\II+f_\II\, . \label{substUTK}
  \eea
If we make the identification
\be
\vartheta_i = \frac{1}{2}\lf(\cot^{-1}\lf[\frac{k }{m_a}\ri] -
\cot^{-1}\lf[\frac{k }{m_i}\ri]\ri), \qquad  (a,i)=(A,{\rm I}),(B,{\rm II})\, .
\ee
the resulting  Hamiltonian \eqref{48a} is now expressed in terms of the $A,B$ modes.

Let us now come back to the full Hamiltonian \eqref{generalH}. After the $V$-limit, in general  we obtain
an Hamiltonian density of the form:
\be\label{HfullV}
  {\widebar {\cal H}}_0 \ = \ \sum_{i={\rm I},{\rm II}}\lf({\cal H}_0^i \ + \ \delta \mathcal{H}_0^i \ \ri)
\ + \ \de {\cal H}_{\rm{mix}}\, .
\ee
In order to select among the inequivalent representations, we have to impose an appropriate renormalization condition on the
form of the Hamiltonian \eqref{HfullV}.

With respect to the simple case, where only one field was present, we have now two distinct possibilities:

The first possibility is to impose the condition that the Hamiltonian \eqref{HfullV} becomes fully diagonal in two fermion
fields, $\psi_1$ and $\psi_2$, with masses $m_1$ and $m_2$:
\be
\label{H12}
{\widebar {\cal H}}_0  \ = \ \sum_{j=1,2}  \widebar{\psi}_j  \left(
-i\vect{\gamma}\cdot\!\vect{\nabla}\ + \  m_j\right)  {{\psi_j}}\, .
\ee
The condition for the complete diagonalization of \eqref{HfullV} is found to be \cite{DynMix}:
\begin{eqnarray}
\theta &\rightarrow& \widebar{\theta} \ \equiv \
\frac{1}{2}\, \tan^{-1}\left[\frac{2h}{m_{\mu}-m_{e}}\right], \label{fullmixdiag-a} \\[2mm]
m_\A & \rightarrow & m_1 \ \equiv \
\frac{1}{2}\left(m_{e}+m_{\mu}-\sqrt{(m_{\mu}-m_{e})^2+4h^2}\right),
\label{fullmixdiag-b}\\[2mm] \label{fullmixdiag}
m_\B & \rightarrow & m_2 \ \equiv \
\frac{1}{2}\left(m_{e}+m_{\mu}+\sqrt{(m_{\mu}-m_{e})^2+4h^2}\right).
\end{eqnarray}
where we introduced the notation $m_{e} = m_\I + f_\I$, $m_{\mu} = m_\II + f_\II$.
The vacuum state associated with such a representation is denoted as
\begin{eqnarray}
|0(\widebar{\theta}, m_1, m_2) \rangle \ \equiv \ |0\ran_{1,2}\, ,
\end{eqnarray}
Another possible representation is obtained by a partial diagonalization
of \eqref{HfullV}, leaving untouched $ \de {\cal H}_{\rm{mix}}$. This will lead to the Hamiltonian density
\be\label{Hem}
{\widebar{\mathcal{H}}}_0  \ = \ \sum_{\si=e,\mu}    \widebar{\psi}_\si \left(
-i\vect{\gamma}\cdot\!\vect{\nabla}\ + \  m_\si\right) {{\psi_\si}}
\,+\,
  h\,  ({\widebar \psi}_e \psi_\mu \ + \ {\widebar \psi}_\mu \psi_{e})\, .
\ee
Such a representation is obtained by setting
\bea
\te &\rar& 0\, ,\\
m_A &\rar & m_e \ \equiv \ m_\I+f_\I\, ,
\\[2mm]
m_B &\rar & m_\mu \ \equiv \ m_\II+f_\II\,.
\eea
The vacuum in this representation is denoted as
\begin{eqnarray}
|0(\theta=0, m_e, m_\mu )\rangle \ \equiv \ |0\rangle_{e\mu}
\, ,
\end{eqnarray}
and will be called the flavor vacuum. We  note  that the mixing term in
Eq.\eqref{Hem} is form-invariant under the transformation
\eqref{4x4Bog}, provided $\te=0$.

Being the representations $|0\ran_{1,2}$ and $| 0 \ran_{e,\mu}$ unitarily inequivalent to each other, it is clear that one
has to make a choice between them based on physical considerations.  In this respect, it seems more reasonable to adopt the representation built on the flavor vacuum  $|0 \ran_{e,\mu}$, as the one which better fits the situation present in the Standard Model, where the flavor fields describe the physical particles, and do not have in general a diagonal mass matrix.

The difference between the two above representations can be also seen via the gap equations which are formally written as
a set of $3$ non-linear equations for $f_\I$, $f_\II$, $h$ and regulate the dynamical generation of both masses and mixing terms. This will be also studied with the effective action approach.

Finally, we note   that the transformation~\eqref{4x4Bog} is of the same form of the one studied in Ref.\cite{BGV}.

\section{Dynamical generation of flavor mixing - Effective action approach}

To study the problem from a functional integral point of view, we start considering the mechanism of dynamical mass generation as done in \cite{Miransky}. The $N$-flavors NJL Lagrangian is written as:
\begin{eqnarray}
\mathcal{L}=i\widebar{\vect{\psi}}\gamma^\mu \partial_\mu \vect{\psi}+G\sum^{N^2-1}_{\alpha=0}\lf[\lf(\widebar{\vect{\psi}}
\frac{\lambda^\alpha}{2}\vect{\psi}\ri)^2+\lf(\widebar{\vect{\psi}} \frac{\lambda^\alpha}{2} i \gamma ^5\vect{\psi}\ri)^2\ri],
\label{lagnjl1}	
\end{eqnarray}
where $\lambda^\alpha$ are the generators of the flavor $U(N)$ group, with the normalization $\mathrm{tr}
\left(\lambda^\alpha \lambda^\bt \right)=2\delta^{\alpha \beta}$. The spinor $\vect{\psi}$ carries a flavor index.

Using the Fierz identity for $\lambda^\alpha$
\begin{eqnarray}
\sum^{N^2-1}_{\alpha=0}\ha \lambda^\al_{ab}\lambda^\al_{cd}=\delta_{ab}\delta_{cd},
\end{eqnarray}
one can rewrite the Lagrangian~\eqref{lagnjl1} in the form:
\begin{eqnarray}
\mathcal{L}=i\widebar{\vect{\psi}}\gamma^\mu \partial_\mu \vect{\psi}+2G\widebar{\psi}^a_L\psi^b_R \widebar{\psi^b}_R\psi^a_L .
\end{eqnarray}
The Lagrangian \eqref{lagnjl1} is invariant under transformations of the chiral group $U_L(N)\times U_R(N)$. We rewrite it as
\be
\mathcal{L}=i\widebar{\vect{\psi}}\gamma^\mu \partial_\mu \vect{\psi}-\widebar{\vect{\psi}}_LM\vect{\psi}_R -\widebar{\vect{\psi}}_RM^\dagger\vect{\psi}_L-\frac{1}{2G}\mathrm{tr}\left(MM^\dagger\right) \label{lagbos},
\ee
where $M$ is an auxiliary boson field that has to respect the following constraint equations:
\be
M_{ab}=-2G\widebar{\vect{\psi}}^b_R \vect{\psi}^a_L\,\,\,\,\,\,\,\,\,\,\,\,\,\,\,\,\,\,\,M^\dagger_{ab}=-2G\widebar{\vect{\psi}}^b_L \vect{\psi}^a_R.
\ee
These are nothing but Euler-Lagrange equations for $M$. The reader can also recognize that the $M$ field is the Hubbard--Stratonovich composite (or effective) field which
can be constructively introduced directly on the level of functional integral via
Hubbard--Stratonovich transformation. It is  interesting to note that, apart the kinetic term of $M$, this is the Lagrangian of the linear $\sigma-$model~\cite{Miransky,Gelevy}.

We write the generating functional $Z[J]$ as
\begin{equation}
					Z[J]=K\int \textit D M \textit D M^\dagger \textit D \psi \textit D\widebar{\psi}\;\exp \left\{i\int \!\! \mathrm{d}^4x \left[\mathcal{L}(x)+F[J]\right]\right\} \, ,
																														\label{gengre}
\end{equation}
where $\mathcal L$ is the Lagrangian density \eqref{lagbos} and $F$ is
\begin{equation}
F[J]=\sum_i J_i(x) \phi_i(x) \, ,
\end{equation}
where $\phi_i$ are all the local fields in the theory, and $J_i$ are the auxiliary currents.	The factor $K$, as usual, is determined so that $Z[0]=1$. Taking into account the explicit form of the Lagrangian we find that
\begin{equation}
Z[J]=K\int \!\!\textit D M \textit D M^\dagger \exp \left\{i\int \!\!\mathrm{d}^4x \left[-\frac{1}{2G}\mathrm{tr}(MM^\dagger)+F[J]\right]\right\}Z_f(M,M^\dagger) \, ,
\end{equation}
where $Z_f$ is
\begin{equation}
Z_f(M,M^\dagger)=\int \textit {D} \psi \textit {D} \widebar{\psi}	\exp\left\{i \int \!\! \mathrm{d}^4x \; \widebar{\psi} \,i \cal{D} \,\psi\right\} \label{functional}
\end{equation}
and ${i \cal{D}}$ is
\begin{equation}
{i \cal{D}}=i\gamma^\mu \partial_\mu- M \otimes\frac{(1-\gamma^5)}{2}-M^\dagger \otimes\frac{(1+\gamma^5)}{2}.
\end{equation}

To evaluate \eqref{functional} we need to perform the Wick rotation and then solve the Gaussian integral thanks to the formula, valid for the Grassman variables:
\begin{equation}
\int \textit {D} \widebar{\psi}\textit {D} \psi \, \exp\left[\int \mathrm{d}^4x \mathrm{d}^4y \, \widebar{\psi}(x)A(x,y)\psi(y)\right]\,=\, C \, \mathrm{Det}{A}	\, ,													\end{equation}
where $C$ is a normalization factor. Thus, it follows that
\begin{equation}
Z_f(M,M^\dagger)\,=\,C \,\mathrm{Det}\, i \cal{D}. 	
\end{equation}
We point out that here the determinant is in a functional sense. The action of the effective theory is then:
\begin{equation}
S(M_c,M^\dagger _c)=i \log (\mathrm{Det}{\,i \cal{D}})+\frac{1}{2G}\int \! \!\mathrm{d}^4 x\, \mathrm{tr}
\left(M^\dagger _c M_c \right).\label{effac}														
\end{equation}
Here $M_c$ plays the role of a classical variable and is defined as:
\begin{equation}
M_c =\frac{\delta W}{\delta J}=\frac{\langle 0|M|0\rangle}{Z[J]},\label{classfield}
\end{equation}
where $W$ is the functional generator of the connected Green's functions.														
Let us note, in fact, that at tree level Eq.\eqref{effac} is the effective action (the generator of amputated Green's functions).

Let us now perform the mean field approximation, i.e. we neglect the fluctuations around the minimum of the potential. Mathematically this means to impose the variational principle
\begin{equation}
	\frac{\delta S}{\delta M_c}=0 .\label{meanfield}
\end{equation}

We search a solution of Eq.\eqref{meanfield} in the form \cite{Miransky}
\begin{equation}
 M_c=\frac{v}{\sqrt{N}}\ide_N\, ,\label{expva}
\end{equation}
where $\ide_N$ is the $N \times N$ identity matrix.
		
To evaluate the first term in Eq.\eqref{effac}, we use the identity	
\begin{equation}
\log \mathrm{Det} A= \mathrm{Tr} \log A, \label{detra}
\end{equation}
where the ``big trace" includes the functional trace and the ``little trace" $\tr A$, that is the traditional matrix trace. Therefore, in our case, we can write:														
\begin{equation}
\log (\mathrm{Det} \, i \mathcal{D})= \mathrm{Tr} \log i \mathcal{D}=
\delta ^4  (0) \int \!\! \mathrm {d}^4 x \, \mathrm{tr}\log \left(\gamma^\mu p_\mu-\frac{v}{\sqrt{N}} \, .\right)
\end{equation}
Here we used that, taking into account Eq.\eqref{expva}:
\begin{equation}
\langle p'|i \mathcal{D}|p \rangle = \delta ^4 \left(p-p'\right)
\left(\gamma^\mu p_\mu-\frac{v}{\sqrt{N}}\right).
\end{equation}
Using the Fourier representation of the Dirac delta, the effective potential is written as
\begin{equation}
V(v)=\frac{2iN}{{(2\pi)}^4}\int d^4p \log \left(1-\frac{v^2}{N p^2}\right) +\frac{1}{2G} v^2 \, .\label{effpo}
\end{equation}
Deriving respect to $v$ and equaling to zero we arrive at the gap equation:
\begin{equation}
m_{dyn}=\frac{4iG}{{(2\pi)}^4}\int \mathrm{d}^4 p \frac{m_{dyn}}{p^2-m^2_{dyn}+i\epsilon} \label{gapd}
\end{equation}
															
Here we put $m_{dyn}=-\frac{v}{\sqrt{N}}$ and we added the poles shift term. This equation has non trivial solutions only if the coupling constant overcome a certain value. When the coupling constant overcomes this limit value, we obtain a tachyonic bosonic bound state and then the vacuum instability. In order to cure this instability the vacuum rearranges itself and gives mass to fermions \cite{Gatto}.											

Let us now try to reformulate, with the effective action formalism, the dynamical generation of fermion mixing, introduced in the previous Section, as done for the dynamical generation of mass.

The general form of M is:
\be
  M=\frac{1}{\sqrt{2}}[(\sigma+i\eta)\ide_2+(\vect{\sigma} +i\vect{\pi}) \cdot \vect{\tau}] \, ,
\ee
where $\ide_2$ is the $2 \times 2$ identity matrix, $\vect{\tau}=(\tau_1,\tau_2, \tau_3)$ is a vector with components the Pauli matrices and $\sigma,\eta,\vect{\sigma},\vect{\pi}$ are a scalar and pseudo-scalar flavor singlet and a scalar and pseudo-scalar flavor triplet, respectively.

We search a solution of Eq.\eqref{meanfield} in a more general form than above:
\begin{equation}
 M_c=
\frac{1}{\sqrt{2}}\begin{bmatrix}
v_0+v_3 & v_1-i v_2 \\ v_1+i v_2 & v_0-v_3
\end{bmatrix} .
		\label{expva2}
\end{equation}
Here we called
\be
 \sigma _c=v_0 \,\,\,\,\,\,\, \vect{\sigma}_c=\mathrm{\G v}=\left\{v_1,v_2,v_3 \right\} \, ,																		\ee
where $\sigma _c$ and $\vect{\sigma}_c$ are the vacuum expectation values of $\sigma$ and $\vect{\sigma}$, respectively, defined as in Eq.\eqref{classfield}. We assumed equal to zero the vacuum expectation values of the pseudo scalar fields.

To evaluate the first term in Eq.\eqref{effac}, we use the identity \eqref{detra}. Thus we can write
\begin{equation}
\log (\mathrm{Det}\, i \mathcal{D})= \mathrm{Tr} \log i \mathcal{D}=  \delta ^4 (0) \int \mathrm {d}^4 p \, \mathrm{tr}\log \left(\gamma^\mu p_\mu-M_c\right),
\end{equation}
and, remembering the explicit form of $M_c$ Eq.\eqref{expva2}:
\begin{equation}
\log (\mathrm{Det} \,i \mathcal{D})= \delta ^4 (0) \int \mathrm {d}^4 p \, \mathrm{tr}\log \left(\gamma^\mu p_\mu-\frac{v_0+\mathrm{\G v}\cdot \vect{\tau}}{\sqrt{2}}\right). \label{eqinacp}
\end{equation}
Note that here is always understood a tensor product and then, this equation, should be of the form:
\be
\log (\mathrm {Det} \,i \mathcal{D})= \delta ^4 (0) \int \mathrm {d}^4 p \, \mathrm{tr}\log \left(\gamma^\mu p_\mu\otimes \ide_2-\frac{v_0 \ide_2+\mathrm{\G v}\cdot \vect{\tau}}{\sqrt{2}}\otimes \ide_4\right).
\ee
Here $\ide_2$ and $\ide_4$, are the $2 \times 2$ and the $4 \times 4$ identity matrices. In the calculations are involved only tensor products between identity matrices and then, to simplify the notation, we drop out these matrices as done in Eq.\eqref{eqinacp}.

The $8 \times 8$ matrix of which we have to evaluate the $\mathrm{tr} \log$ is
\be
A_8=\begin{bmatrix}
 \ide_4 \left(1-\frac{{v_0}^2+{{|\mathrm{\G v}|}^2}}{2 p^2}-\frac{{v_0} {v_3}}{p^2} \right)& -\ide_4 \left(\frac{{v_0} ({v_1}-i {v_2})}{p^2} \right)\\
 -\ide_4\left(\frac{{v_0} ({v_1}+i {v_2})}{p^2}\right) & \ide_4\left(1-\frac{{v_0}^2+{{|\mathrm{\G v}|}^2}}{2 p^2}+\frac{{v_0} {v_3}}{p^2}\right)
\end{bmatrix} \,\,.\label{8matrix}
\ee																								
Thus the effective potential is:
\begin{equation}
V(v_0,v_1,v_2,v_3)=\frac{{v_0}^2+{|\mathrm{\G v}|}^2}{2 G}+\frac{i}{8 \pi ^4} \int \mathrm{d}^4 p \log \left(\frac{4 p^4-4 p^2 \left({v_0}^2+{|\mathrm{\G v}|}^2\right)+\left(-{v_0}^2+{|\mathrm{\G v}|}^2\right)^2}{4 p^4}\right)
\end{equation}
We note that this reduces to \eqref{effpo} when $|\vect{v}|=0$. Deriving respect to $v_0,v_1,v_2,v_3$ and equaling the result to zero we arrive at the explicit form of the gap equations:
\bea
\frac{{v_0}}{G}&=& -\frac{i}{8 \pi ^4}\int \mathrm{d}^4 p\frac{\left[-8 p^2 {v_0}-4 {v_0} \left(-{v_0}^2+{|\mathrm{\G v}|}^2\right)\right]}{\left[4 p^4-4 p^2 \left({v_0}^2+{|\mathrm{\G v}|}^2\right)+\left(-{v_0}^2+{|\mathrm{\G v}|}^2\right)^2\right]} \label{gap0}
\\
\frac{{\mathrm{\G v}}}{G}&=&-\frac{i}{8 \pi ^4} \int \mathrm{d}^4 p\frac{\left[4 {\mathrm{\G v}} \left(-{v_0}^2+{|\mathrm{\G v}|}^2\right)-8 p^2 {\mathrm{\G v}}\right]}{\left[4 p^4-4 p^2 \left({v_0}^2+{|\mathrm{\G v}|}^2\right)+\left(-{v_0}^2+{|\mathrm{\G v}|}^2\right)^2\right]} \label{gapv}
\eea
In parallel with the treatment of the previous Section we can consider two cases:
\medskip

\noi {1.} \hspace{.2cm}
One can search a solution in the form:
\begin{equation}
M_c=
\frac{1}{\sqrt{2}}\begin{bmatrix}
v_0+v_3 & 0 \\ 0 & v_0-v_3
\end{bmatrix}.
\label{expva3}
\end{equation}
Calling $m_A=\frac{1}{\sqrt{2}}(v_0+v_3)$ and $m_B=\frac{1}{\sqrt{2}}(v_0-v_3)$ and substituting $M_c$ in the Lagrangian \eqref{lagnjl1}, we obtain:
\be
\mathcal{L}=\sum_{a=A,B}\left(i\widebar{\psi}_a\gamma^\mu \partial_\mu \psi_a-m_a \widebar{\psi}_a\psi_a\right) \label{lagdiag},
\ee
that corresponds to the case of \eqref{48a}.

\medskip

\noi {2.} \hspace{.2cm}
More in general we can also search a solution in the form:
\begin{equation}
M_c=\frac{1}{\sqrt{2}}
\begin{bmatrix}
	v_0+v_3 & v_1 \\ v_1 & v_0-v_3
\end{bmatrix}.
\label{expva4}
\end{equation}
Putting $f_\mathrm{I}=\frac{1}{\sqrt{2}}(v_0+v_3)$, $f_\mathrm{II}=\frac{1}{\sqrt{2}}(v_0-v_3)$ and $h=\frac{v_1}{\sqrt{2}}$ and substituting in the Lagrangian \eqref{lagnjl1} we obtain:
\be
\mathcal{L}=\sum_{a=\mathrm{I},\mathrm{II}}\left(i\widebar{\psi}_a\gamma^\mu \partial_\mu \psi_a-f_a \widebar{\psi}_a\psi_a\right)-h\left(\widebar{\psi}_\mathrm{I}\psi_\mathrm{II}+\widebar{\psi}_\mathrm{II}\psi_\mathrm{I}\right) \label{lagpardiag},
\ee
that corresponds to the case of \eqref{HfullV}.
At this point we consider two possibilities:
\smallskip

\noi $\bullet$ We can diagonalize completely the Lagrangian \eqref{lagpardiag}. To do this we have to diagonalize the mass matrix \eqref{expva4}. From the secular equation we obtain, calling the eigenvalues $m_1$ and $m_2$:
\begin{eqnarray}
 m_1 \ =\
\frac{1}{2}\left(f_\mathrm{I}+f_\mathrm{II}-\sqrt{(f_\mathrm{II}-f_\mathrm{I})^2+4h^2}\right),
\label{fullmixdiag-b1}\\[2mm] \label{fullmixdiag1}
m_2 \ = \
\frac{1}{2}\left(f_\mathrm{I}+f_\mathrm{II}+\sqrt{(f_\mathrm{II}-f_\mathrm{I})^2+4h^2}\right),
\end{eqnarray}
that are Eqs.\eqref{fullmixdiag-b}-\eqref{fullmixdiag} in the case in which $m_\mathrm{I}$ and $m_\mathrm{II}$ are zero. Let us now search the eigenvectors of the matrix \eqref{expva4}. The eigenvalue equation is:
					\be
\begin{bmatrix}
f_\mathrm{I} & h \\ h & f_\mathrm{II}
\end{bmatrix}
\begin{bmatrix}
x_i \\ y_i
\end{bmatrix} = m_i
\begin{bmatrix}
 x_i \\ y_i
\end{bmatrix}\,\,\,\,\,\,\,\,\, i=1,2. \label{eigeq}
\ee
Is well known that $M_c$, being symmetric can be diagonalized by an orthogonal matrix $G$, that has the eigenvectors as columns:
\be
G=  \begin{bmatrix}
x_1 & x_2 \\ y_1 & y_2
\end{bmatrix}.
\ee
The relation between $M_c$ and the diagonal matrix $\tilde{M}$ is:
\be
			\tilde{M}=G^{-1}M_c G  \label{simtra}
\ee
From Eq.\eqref{eigeq} we find that $G$ takes the form
\be
		  G=  \begin{bmatrix}
					x_1 & x_2 \\ \frac{m_1-f_I}{h}x_1 & \frac{m_2-f_I}{h}x_2 \label{diagma}
				 \end{bmatrix} \,\, .
		\ee
We have the freedom to choose $x_1$ and $x_2$. Our choice is to have diagonal elements equal to one. We obtain:
\be
G= \begin{bmatrix}
				1 & \frac{2h}{\left[1+\sqrt{1+({\frac{2h}{f_\mathrm{II}-f_\mathrm{I}})}^2}\right]
(f_\mathrm{II}-f_\mathrm{I})} \\ -\frac{2h}{\left[1+\sqrt{1+({\frac{2h}{f_\mathrm{II}-f_\mathrm{I}})}^2}\right]
(f_\mathrm{II}-f_\mathrm{I})} & 1 \label{diagma3}
	 \end{bmatrix}.
\ee
If we call $\zeta=\frac{2h}{\left[1+\sqrt{1+({\frac{2h}{f_\mathrm{II}-f_\mathrm{I}})}^2}\right](f_\mathrm{II}-f_\mathrm{I})}$ and we remember that the matrix $G$ is defined up to a constant we reach at the following form:
\be
G=\frac{1}{\sqrt{1+\zeta^2}} \bbm
 1 & \zeta \\ -\zeta & 1 															                       \ebm  \label{matcoh} \, .
\ee
Let us note that the matrix $G$ depends only on one parameter $\tan (2\theta)\equiv\frac{2h}{f_\mathrm{II}-f_\mathrm{I}}$. This is the same parameter that appears in Eq.\eqref{fullmixdiag-a}. This result was now derived in an independent way. Moreover this matrix belongs to $SU(2)/U(1)$ and then has the form of a generator of generalized coherent states \cite{Perelomov}. The Lagrangian \eqref{lagnjl1}, in the mass basis, is then written as:
\be
 \mathcal{L}=\sum_{a=1,2}\left(i\widebar{\psi}_a\gamma^\alpha \partial_\alpha \psi_a-m_a \widebar{\psi}_a\psi_a\right) \label{lagdiagnt}
\ee
It is very important to note that this case is substantially different respect to the case of Eq.\eqref{lagdiag}. Now there are three bosons in the vacuum, while in the previous case there were only two bosons. This is the physical nature of the inequivalence between these two situations.

An helpful relation can be found between $\theta$ and $\zeta$:
\be
			      \zeta=\tan \theta \, .
\ee
This relation with Eq.\eqref{simtra} leads to Pontecorvo mixing formula.
								
\smallskip

\noi $\bullet$ We suppose that all the possible physical configuration can be obtained by $M_c$ through a similarity transformation like in Eq.\eqref{simtra}, with $G$ of the form \eqref{matcoh}. In the case of $\zeta=0$, putting $f_\mathrm{I}=m_e$ and $f_\mathrm{II}=m_\mu$, we find, substituting in the Lagrangian \eqref{lagnjl1}:
\be
   \mathcal{L}=\sum_{a=\mathrm{e},\mathrm{\mu}}\left(i\widebar{\psi}_a\gamma^\alpha \partial_\alpha \psi_a-m_a \widebar{\psi}_a\psi_a\right)-h\left(\widebar{\psi}_\mu\psi_e+\widebar{\psi}_e\psi_\mu\right) \label{lagpardiag1}.
			\ee
Then we recovered the situation of Eq.\eqref{Hem}.

Therefore thanks to the language of the effective action, in the classical, mean field limit, we have recovered all the previous cases. Moreover we found an interesting connection: every inequivalent representation (physical phase of the system) can be put in connection with an $SU(2)$ coherent states. The space of all generators of $SU(2)$ coherent states, as known \cite{Perelomov}, has the structure of a K\"ahlerian manifold. In particular the case of the flavor vacuum coincides with the minimum of the K\"ahlerian potential
\be
	  F=\log{(1+{|\zeta|}^2)} \,.
\ee
\section{Conclusions}

In this paper we have considered the problem of dynamical generation of flavor mixing in the context of Nambu-Jona Lasinio model for the simplest case of two generations of Dirac fields. We have first
reviewed an operatorial approach to this problem already presented in Ref.\cite{DynMix} and based on the inequivalent representations in the spirit of  an early treatment given in Ref.\cite{UTK} for the case of one generation (dynamical mass generation). We found that the patterns of dynamical symmetry breaking and the related vacuum structures are essentially different in the two cases when mixing is present or not at physical level, although at operatorial level the respective Hamiltonians are connected only by a rotation in the fields.

We have then considered the same problem in the functional integral formalism, by studying one-loop effective action and obtaining gap equations which include also the dynamical generation of mixing terms. The preliminary results here presented seem to confirm what found in the operatorial formalism, although more study is necessary to fully connect the two treatments.
In this respect, it appears very interesting the general issue about the capability of functional formalism to take into account inequivalent representations and possibly its formal extension in this sense.

\acknowledgments
P.J. was supported by the GA\v{C}R Grant No. GA14-07983S.


\end{document}